\documentclass[fleqn,usenatbib]{mnras}

\usepackage{newtxtext,newtxmath}

\usepackage[T1]{fontenc}
\usepackage{ae,aecompl}

\usepackage{graphicx}	
\usepackage{amsmath}	
\usepackage{gensymb}
\usepackage{verbatim}

\title[V404 Cygni with \textit{Fermi}-LAT]{V404 Cygni with \textit{Fermi}-LAT}

\author[M. Harvey et al.]{
Max Harvey,$^{1}$\thanks{E-mail: max.harvey@durham.ac.uk}
Cameron B. Rulten,$^{1}$
and Paula M. Chadwick$^{1}$
\\
$^{1}$Centre for Advanced Instrumentation, Department of Physics, University of Durham, South Road, Durham DH1 3LE, United Kingdom\\
}

\date{Accepted XXX. Received YYY; in original form ZZZ}

\pubyear{2019}

\begin{document}
\label{firstpage}
\pagerange{\pageref{firstpage}--\pageref{lastpage}}
\maketitle

\begin{abstract}
We revisit the well-studied outburst of the low mass X-ray binary (LMXB) system V404\,Cygni, and claims of $\gamma$-ray excesses observed with the \textit{Fermi}-LAT instrument. Upon analysing an 11.5 year dataset with the 8-year LAT point source catalogue and 8-year background models, we find no evidence to suggest that there is high energy $\gamma$-ray emission during the outburst period (or at any other time) from V404\,Cygni. This is due to the proximity of V404\,Cygni to the $\gamma$-ray emitting blazar B2023+336, a luminous source approximately $0.3 \degree$ away, which causes source confusion at the position of V404\,Cygni, the luminous $\gamma$-ray background, and the use of older background models and catalogues in previous studies.
\end{abstract}




\section{Introduction}
\label{intro}

\subsection{X-ray binaries and $\gamma$-ray emission}
V404\,Cygni (also known as GS\,2023+338) is an X-ray binary system which lies on the Galactic plane \citep{kitamoto_gs2023_1989}. It consists of a K-type star with a mass slightly below $1 M_{\odot}$ \citep{wagner_periodic_1992} and a black hole companion of mass $9 M_{\odot}$ \citep{khargharia_near-infrared_2010}; this stellar mass means that V404\,Cygni is considered a low mass X-ray binary (LMXB). This system is visible across the EM spectrum, with X-ray emission originating from accretion of the donor star onto its black hole companion; in common with other LMXBs, the accretion occurs through overflow of the star's Roche lobe onto the black hole. 

High energy $\gamma$-ray emission has been observed from a variety of binary systems, with the emission thought to originate in one of three mechanisms. In the first, the spin-down of young pulsars causes $\gamma$-ray emission through relativistic shocks \citep{dubus_gamma-ray_2006}, these systems are typically referred to as \textit{$\gamma$-ray binaries}. The second type of system is a colliding wind binary, where interactions between the stellar winds of two massive stars produce $\gamma$-rays up to TeV energies. The only colliding wind binary detected with \textit{Fermi}-LAT is $\eta$-Carinae \citep{leser_first_2017}. Finally, a class of binaries known as the microquasars is predicted to produce $\gamma$-rays through the acceleration of particles in jets (\citealt{bosch-ramon_broadband_2006}, \citealt{orellana_leptonic_2007}, \citealt{araudo_high-energy_2009}). Jets are seen from microquasars in a variety of states  as described by the hardness-intensity cycle \citep{fender_towards_2004}, including compact jets in the low-hard X-ray state \citep{tomsick_watching_2008}. It is from these jets that the particle acceleration required for $\gamma$-ray emission takes place, with evidence to suggest that $\gamma$-rays are produced when a jet forms at the transition between the hard and soft state, and also from the hard state compact jet \citep{zdziarski_high-energy_2017}.

There are a number $\gamma$-ray emitting microquasar objects known. The first and arguably best studied is SS\,433, which is a unique case as it is the only X-ray binary known to accrete in a constant supercritical way \citep{fabrika_jets_2004}.  Other well-known $\gamma$-ray emitting microquasars more comparable to V404\,Cygni include Cygnus\,X-3 \citep{tavani_extreme_2009} (where orbital $\gamma$-ray modulation is seen \citep{abdo_modulated_2009}) and Cygnus\,X-1, where $\gamma$-ray emission is seen during the hard spectral state (\citealt{bodaghee_gamma-ray_2013}, \citealt{zanin_gamma_2016}, \citealt{zdziarski_high-energy_2017}). As V404 Cygni is not in a supercritical accretion regime, it is thought to produce $\gamma$-ray emission through entering an outburst like Cygnus X-1 and Cygnus X-3.

\subsection{Outbursts of V404\,Cygni}

There have been at least 3 historic outbursts of V404\,Cygni during the 20th Century: one in 1938 (when the system was designated Nova\,Cygni\,1938), one in 1958 (undetected at the time), a possible outburst in 1979, and the well-studied outburst of 1989 (\citealt{duerbeck_reference_1988}, \citealt{richter_v404_1989}, \citealt{wagner_v404_1989}, \citealt{makino_gs_1989} and \citealt{han_radio_1992}). Together with the 2015 outburst, this means that V404 Cygni has an outburst approximately every few decades. Prior to the 1989 outburst, these events were recorded only in the optical waveband; the 1989 outburst was also captured in the X-ray waveband by the \textit{Ginga} X-ray Astronomy Satellite and the system was designated GS\,2023+338 \citep{kitamoto_gs2023_1989}. The simultaneous X-ray and optical activity of the 1989 outburst led to the rapid association of GS\,2023+338 with V404\,Cygni \citep{wagner_v404_1989}.

In June 2015 \textit{Swift}-BAT reported that V404\,Cygni had begun an outburst with enhanced fluxes reported across the spectrum, with \textit{Fermi}-GBM triggering approximately 30 minutes later.  At this time, an enhancement in accretion rate caused a subsequent enhancement in jet brightness, sending V404\,Cygni into an outburst. Before it returned to its pre-outburst flux level in August 2015, INTEGRAL, \textit{Swift}, AGILE, MAGIC, and VERITAS all took observations at high energies, with \textit{Fermi}-LAT's all sky coverage also capturing the position of V404\,Cygni during its outburst. There was also a brief `sequel' flare in December 2015, lasting for approximately 2 weeks. However, for the purposes of this paper we consider the outburst as exclusively within the June 2015 - August 2015 period.

The \textit{Swift} team reported a variable X-ray flux, at times in excess of 40 times the Crab Nebula flux (\citealt{barthelmy_swift_2015}, \citealt{motta_swift_2017}), in addition to enhanced and variable UV and optical fluxes \citep{oates_swift_2019}. Enhanced fluxes from the V-band optical up to the soft $\gamma$-rays (40 - 100 keV) were observed with INTEGRAL (\citealt{rodriguez_correlated_2015} \& \citealt{roques_first_2015}). The AGILE (50 - 400 MeV $\gamma$-ray) team reported a $4.3 \sigma$ enhancement of $\gamma$-ray flux, contemporaneous with a large flare in the radio and X-ray wavebands between MJD 57197.25 - 57199.25, though no significant $\gamma$-ray emission above 400 MeV was reported \citep{piano_high-energy_2017}, with a similar excess also being observed with \textit{Fermi}-LAT \citep{loh_high-energy_2016}. In the very high energy (GeV to TeV) range, the VERITAS \citep{archer_very-high-energy_2016} and MAGIC collaborations \citep{ahnen_magic_2017} both reported upper limits from the position of V404\,Cygni. 

\section{V404 Cygni as seen with \textit{Fermi}-LAT}

The \textit{Fermi}-LAT has an effective energy range of 100 MeV to 300 GeV, which essentially bridges the gap in energy between AGILE (operating in the MeV range) and Cherenkov telescope systems like VERITAS and MAGIC (operating in the GeV-TeV range). Although a $4.3 \sigma$ enhanced $\gamma$-ray flux was below 400 MeV was seen with AGILE, there was no significant detection over the 10 hour exposure of MAGIC (VERITAS had a shorter exposure of 2.5 hours). It is questionable whether one would expect to see an enhanced flux with \textit{Fermi}-LAT.

At the position of V404\,Cygni, there is no catalogued source in the most recent \textit{Fermi}-LAT point source catalogue, the 4FGL \citep{abdollahi_fermi_2020}. This is to be expected, as the 4FGL uses an 8 year observation time to detect sources, whereas we might expect high energy $\gamma$-rays to be produced only when V404\,Cygni is in an outburst. Even if V404\,Cygni was luminous in $\gamma$-rays throughout its outburst, the long observation time would render this emission insignificant. 

However, the 4FGL does catalogue a luminous $\gamma$-ray emitting flat spectrum radio quasar (FSRQ), B2023+336, approximately $0.3 \degree$ away from the position of V404\,Cygni, detected through the Galactic plane \cite{kara_gamma-ray_2012}. This is problematic, as the resolution of the LAT varies from between an optimal $0.15 \degree$ at $>10 \; \mathrm{GeV}$ down to a substantially poorer resolution of $3.5 \degree$ at $100 \; \mathrm{MeV}$ (see Figure 17 in \citealt{atwood_large_2009}). As a result, source confusion between B2023+336 and V404\,Cygni becomes the primary issue in reliably detecting $\gamma$-ray emission from the LMXB with \textit{Fermi}-LAT at any but the highest energies. In addition, V404\,Cygni is located on the Galactic plane, a significant steady source of background photons, which are non-trivial to model, although the most recent Galactic diffuse model improves on previous releases \citep{abdollahi_fermi_2020}. This presents additional challenges to resolving any $\gamma$-ray emission from V404\,Cygni.

\subsection{The results of Loh et al. and Piano et al.}

\cite{loh_high-energy_2016} (referred to as \textit{Loh\,16} throughout this text) explore the \textit{Fermi}-LAT data in the period around the outburst of V404\,Cygni, performing a comprehensive variability analysis at the position of V404\,Cygni. They use photons across most of the \textit{Fermi}-LAT energy spectrum ($100 \; \mathrm{MeV}$ to $100 \; \mathrm{GeV}$), but discard the quartile of photons with the poorest point spread function (PSF) label. While this allows for the better localisation of remaining $\gamma$-ray emission in a model, significant cuts to the number of photons make it more difficult to detect sources, particularly faint and transient ones such V404\,Cygni. 

\textit{Loh\,16} perform a variability analysis on the position of V404\,Cygni by first carrying out a binned analysis of the region and then executing an unbinned light-curve time-series analysis. The bins used are 12 hours (shifted by 6 hours), and 6 hours (shifted by 1 hour) in duration. Based on these results, an excess in the $\gamma$-ray flux is found near the position of V404\,Cygni with a peak test statistic (TS) of 15.3 in one particular 6 hour bin at approximately MJD 57199. The test statistic of the 12 hour bin containing this 6 hour bin is approximately 11. The authors describe this transient excess of having a chance probability of occurring as $2\%$ (giving a \textit{z}-score of approximately $2 \sigma$) based on 320 trials. They state that this gives a $4 \times 10^{-4}$ chance probability of occurring at the same time as a peak in the \textit{Swift}-BAT flux light-curve. For the purposes of this paper, we will describe this event as the June 2015 excess.

The analysis of \textit{Loh\,16} used the most up-to-date models and catalogues for the \textit{Fermi}-LAT data at the time: the \texttt{gll\_iem\_v06} Galactic background model, the \texttt{iso\_P8R2\_SOURCE\_V6\_v06} isotropic background model, and the 3FGL catalogue and extended source templates. Improved background models, the 4FGL catalogue, and improved instrument response functions for the LAT are now available. These allow for an improved analysis of the \textit{Fermi}-LAT data at the time and position of the V404\,Cygni outburst. 

\textit{Loh\,16} is not the only paper discussing the June 2015 excess as seen with \textit{Fermi}-LAT. \cite{piano_high-energy_2017} (hereafter \textit{Piano\,17}), while discussing this same event as seen with the AGILE $\gamma$-ray Imaging Detector (GRID), provide an independent \textit{Fermi}-LAT analysis which complements both the results from the AGILE telescope and those of \textit{Loh\,16}. In the analysis of the AGILE data, \textit{Piano\,17} consider photons in two energy bands: $50 \, \mathrm{MeV}$ - $400 \, \mathrm{MeV}$ and $400 \, \mathrm{MeV}$ - $30 \, \mathrm{GeV}$. \textit{Piano\,17} record a $\gamma$-ray excess of $TS = 18.1$, ($4.3 \sigma$ for 1 degree of statistical freedom) in this first energy band, at a time coincident with the excess recorded by \textit{Loh\,16}. There is no detection at energies $> \, 400 \, \mathrm{MeV}$. 

In their \textit{Fermi}-LAT analysis \textit{Piano\,17} use a different set of photon cuts to \textit{Loh\,16}. They analyse photons across all 4 PSF quartiles, using the \texttt{P8R2\_TRANSIENT\_v16} photon class with the same background models as \textit{Loh\,16} (the \texttt{gll\_iem\_v06} Galactic model and \texttt{iso\_P8R2\_SOURCE\_V6\_v06} isotropic model). \textit{Piano\,17} report a TS of 13.4 in the 24 hour period from MJD 57198.75-57199.75, temporally coincident with the result of \textit{Loh\,16}.

Each \textit{Fermi}-LAT photon class has a corresponding isotropic model and instrument response function, and it is good practice when carrying out LAT data analysis to use these together. \textit{Piano\,17} use the \texttt{P8R2\_TRANSIENT\_v16} photon class with the \texttt{iso\_P8R2\_SOURCE\_V6\_v06} isotropic model, and an unspecified instrument response function (IRF). The use of a mismatched isotropic background and photon class will result in systematic errors in source analysis, and in the misidentification of cosmic rays, reducing the accuracy of \textit{Piano\,17}'s results, though without knowing the IRF it is difficult to assess by how much. Consequently we primarily deal with the LAT results from \textit{Loh\,16} in this paper, as they do not have this additional uncertainty, and find more significant $\gamma$-ray emission at the LAT data from the position of V404\,Cygni.

\subsection{The results of Xing and Wang}
In a more recent paper, \cite{xing_detection_2020} (henceforth referred to as \textit{Xing\,20}), carried out an an analysis of the \textit{Fermi}-LAT data, independent to that of \textit{Loh\,16}. \textit{Xing\,20} employs the most recent 4FGL catalogue and corresponding background models to perform a variability analysis over the mission time of the \textit{Fermi} satellite. \footnote{It should be noted that at the time of publication, \textit{Xing\,20} is available only at arxiv.org, and has not been published in a peer reviewed journal.}

In order to test the results of \textit{Loh\,16}, the authors repeat their variability analysis using a similar overlapping time binning scheme at the same time, using a binned analysis. \textit{Xing\,20} do not state which event class and event type were used in their LAT data analysis. They find no significant $\gamma$-ray flux at the peak of the \textit{Swift}-BAT X-ray flux (on MJD 57199), which is the result of \textit{Loh\,16}. \textit{Xing\,20} suggests that the June 2015 $\gamma$-ray excess of \textit{Loh\,16} was a result of employing older (and poorer) catalogue, background models and IRF, rather than a genuine $\gamma$-ray flare. The lack of detail concerning the analysis parameters and the lack of TS maps below 300\,MeV (where the flux of the $\gamma$-ray emission reported by \textit{Loh\,16} is highest) make it very difficult to reproduce the results of \textit{Xing\,20}.

\textit{Xing\,20} do claim a separate $\gamma$-ray excess ($\mathrm{TS \approx 15}$) during August 2015, which is towards the end of the V404\,Cygni outburst (henceforth referred to as the August 2015 excess). This analysis does not account for any photons with an energy less than $300 \, \mathrm{MeV}$, in contrast to the \textit{Xing\,20} analysis of the June 2015 excess. More significantly, they claim detection of $\gamma$-ray emission from V404\,Cygni at the $7\sigma$ level in August 2016 (the August 2016 excess). This is approximately a year after the June 2015 outburst finishes, and \textit{Xing\,20} report that there is no corresponding increase in X-ray flux in the \textit{Swift}-BAT light-curve at this time.

In this paper, we provide an independent analysis of the reported V404\,Cygni $\gamma$-ray excesses, and of the nearby blazar B2023+336. We investigate the hypothesis of source confusion between this blazar, known to have soft $\gamma$-ray emission, and the position of V404\,Cygni.

\section{\textit{Fermi}-LAT Observations and Analysis.}
\label{obs}

The goal of our analysis is to detect any $\gamma$-ray emission from V404\,Cygni during both its 2015 outburst and August 2016, when \textit{Xing\,20} claim detection of a $\gamma$-ray flare. We take 11.5 years of \textit{Fermi}-LAT data (inclusive of 2015-2016) with photons across the full effective energy range of the instrument: $100 \, \mathrm{MeV}$ to $300 \, \mathrm{GeV}$.

We follow a standard data reduction chain using the \textit{Fermitools} (\texttt{v1.2.23}) and Pass 8 \textit{Fermi}-LAT data, which has improved analysis methods and event reconstruction over previous versions. Following the method of \cite{mattox_likelihood_1996}, we execute a standard binned likelihood analysis using the parameters described in Table \ref{tbl:params}.  Our binned analysis employs $0.1 \degree$ spatial bins in RA and Dec (approximately the optimal resolution of the LAT at high energies), and we bin in energy with 8 bins per decade. Although unbinned analysis is typically used for time series analysis of \textit{Fermi}-LAT sources on short timescales, we employ a binned analysis (as recommended in the Fermitools Cicerone) because V404\,Cygni is on the Galactic plane.

We then follow the method of \cite{mattox_likelihood_1996} to use maximum likelihood estimation to fit a model to the data-set on a bin-by-bin basis. We use the 4FGL catalogue and background parameters described in Table \ref{tbl:params} to make a prediction for the number of photons per bin, and then iteratively push the parameters in the model closer to their likely values in order to improve our model's accuracy.

\begin{table}
\centering
\begin{tabular}{cc}
\hline \hline
Observation Period (Dates) & 04/08/2008 - 10/01/2020 \\
Observation Period (MET) & 239557417 - 600307205 \\
Observation Period (MJD) & 54682 - 58423 \\
Energy Range (GeV) & 0.1 - 300 \\
evtype & 3  (FRONT + BACK) \\
evclass & 128 (\texttt{P8R3\_SOURCE}) \\
Data ROI width & $25\degree$ \\
Model ROI Width & $30\degree$ \\
Zenith Angle & $< 90\degree$ \\
Instrument Response  & \texttt{P8R3\_SOURCE\_V2} \\
Isotropic Background Model & \texttt{iso\_P8R3\_SOURCE\_V2\_v1} \\
Galactic Background Model & \texttt{gll\_iem\_v07} \\
Point Source Catalogue & 4FGL \\

\hline
\end{tabular}
\caption{The parameters used in the likelihood analysis of the region of interest around V404 Cygni.}
\label{tbl:params}
\end{table}

We next free the normalization of all point sources within $5\degree$ of the central position of the region of interest (ROI) as well as the normalization of the Galactic and isotropic diffuse backgrounds. We then perform a full likelihood fit with respect to our freed sources and backgrounds. 

We next employ some of the advanced analysis tools from \texttt{Fermipy} (\texttt{v0.19.0}) \citep{wood_fermipy:_2017}, a \texttt{Python} module which acts as a wrapper for the Science Tools, to further investigate the ROI. We first search for uncatalogued point sources using the \texttt{Find Sources} algorithm, which fits a point source to each spatial bin in the model then calculates a likelihood test statistic for that point source. The test statistic (TS) is defined as the ratio of the likelihood of an alternative ($\Theta_{1}$) and a null hypothesis ($\Theta_{2}$); given by Equation \ref{eqn:ts}.

\begin{equation}
    \label{eqn:ts}
    \mathrm{TS} = 2 \ln{\frac{L(\Theta_{1})}{L(\Theta_{2})}}
\end{equation}

In this case, the null hypothesis is that there is no point source at a position, and the alternative hypothesis is that there is one. Through Wilks' Theorem \citep{wilks_large-sample_1938} the TS equates to a $\chi^{2}$ statistic for $k$ degrees of freedom.

Using this algorithm, we are able to search for new sources in an unbiased way. Whilst we do not expect any long-term $\gamma$-ray emission from V404\,Cygni, any point sources nearby that are uncatalogued will be added to the model, improving its accuracy. We set the TS threshold for the addition of a new point source to our model as $9$ (a $z$-score of $3 \sigma$).

Given the proximity of B2023+336 to V404\,Cygni in the sky, it is important to understand the characteristics of the blazar. Extended $\gamma$-ray emission is detected with \textit{Fermi}-LAT in only two AGN: Fornax\,A \cite{ackermann_fermi_2016} and Centaurus\,A \cite{abdo_fermi_2010}, both radio galaxies. We do not expect to see any extended emission from B2023+336, as it is a blazar, all of which are point sources. We use Equation \ref{eqn:ts} to fit a radial Gaussian model as an alternative hypothesis against the null hypothesis of a point source model. The TS of extension for  B2023+336 is calculated to be -0.01, strongly favouring the point source model over an extended one, as we would expect. 

\subsection{Spectral analysis}
\subsubsection{Spectral analysis of B2023+336}
\label{blz_spectra}
B2023+336 is a notable $\gamma$-ray blazar, as it is one of the few that is seen through the Galactic plane, which is itself a luminous $\gamma$-ray emitter. As a result, although we expect some contamination of the photons from B2023+336 with those from the Galactic plane, which has a soft spectrum, use of the 8 year Galactic background model should minimise this. In this section, we detail our spectral analysis of B2023+336, and compare it with the spectral analyses of the three $\gamma$-ray excesses. If the spectrum of any excess is significantly different from that of the blazar, we can consider this evidence that the excess is not a product of source confusion between B2023+336 and the position of V404\,Cygni. If the spectra are similar this may imply source confusion, but this could be coincidental. Further evidence, such as correlated variability between B2023+336 and V404\,Cygni, would be needed to draw a firm conclusion.

\begin{figure}
    \centering
    \includegraphics[width=240pt]{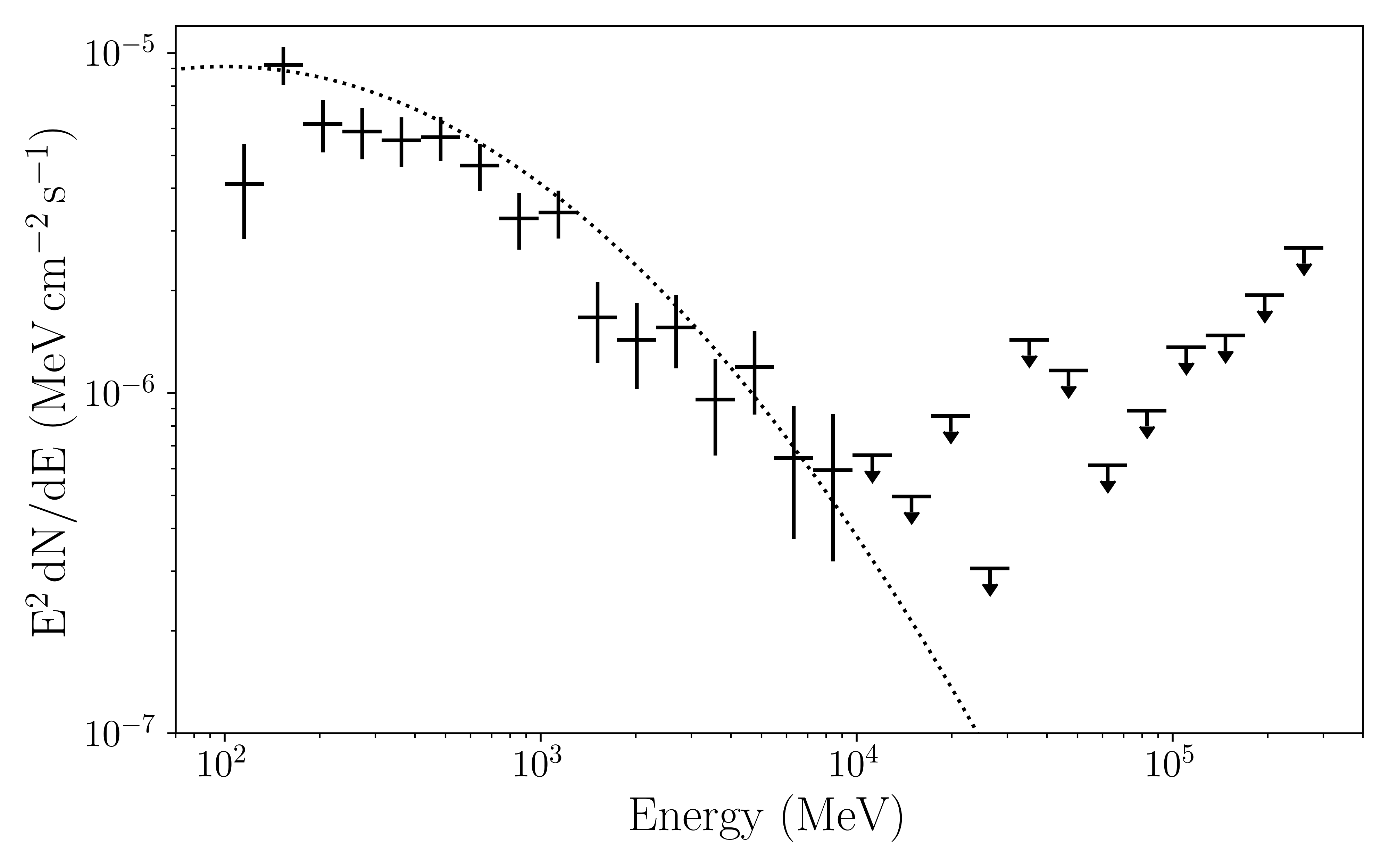}
    \caption{The Spectral Energy Distribution of the blazar, B2023+336, with $\mathrm{E^{2}\frac{dN}{dE}}$ flux shown plotted against bin energy. The dotted line shows our log-parabola fit, using the parameters described in Section \ref{blz_spectra}, with a good fit to the data. This is unsurprising, as a log-parabola spectral shape is common amongst the LAT detected blazar population. We regard any bin which does not have a TS value of at least 4 as an upper limit. }
    \label{fig:sed}
\end{figure}

The spectral energy distribution of the blazar B2023+336 is given in Figure \ref{fig:sed}, where $95\%$ confidence upper limits are fixed to bins with a $z$-score of less than 2. We find that the best fit to the bins is a log-parabola spectral shape, with a $z$-score of $4.4 \sigma$ against a power law model. We also note that the power law with an exponential cutoff (PLEC) model fits to a $4.0 \sigma$ significance, so the difference in the goodness of fit between the log-parabola and PLEC is marginal. For the photon index of the SED, we use the log-parabola index, $\Gamma_{\mathrm{LP}} = 2.74 \pm 0.08$, which is in reasonable agreement with the photon index if a power law was fitted instead ($\Gamma_{\mathrm{PL}} = 2.65 \pm 0.05$). For the power law with exponential cutoff we have a slightly lower photon index ($\Gamma_{\mathrm{PLEC}} = 2.20 \pm 0.05$) with an exponent index of $0.66 \pm 0.19$. Both the log-parabola and PLEC models provide similarly good fits over a power law, and all models describe the spectral shape in a similar way: flux generally anti-correlated with energy, commonly known as a soft $\gamma$-ray spectrum. There is a peak flux of $9.22 \times 10^{-6} \, \mathrm{MeV \, cm^{-2} \, s^{-1}}$ in the energy range $133 \,  \mathrm{MeV}$ to $177 \, \mathrm{MeV}$. We see a $\gamma$-ray flux cut-off at energies above $9.7 \, \mathrm{GeV}$. This spectral fit is compatible with that described in the 4FGL-DR2 \citep{abdollahi_fermi_2020}, where the log-parabola index is given as $\Gamma_{\mathrm{LP}} = 2.73 \pm 0.07$. We also find no evidence for detectable variability of the spectral shape of B2023+336, leading us to believe that our best fit spectral parameters are an accurate description at all times.

\subsubsection{Spectral analysis of the V404\,Cygni $\gamma$-ray excesses}
\label{sec:spectra}
The most significant of the three excesses reported in \textit{Loh\,16} and \textit{Xing\,20} is the August 2016 excess. The authors fit a power law spectral model to this excess, although there are only 4 energy bins significant enough to be plotted. Their spectral fit has a power-law photon index of $\Gamma_{\mathrm{PL}} = 2.9 \pm 0.3$, indicating soft $\gamma$-ray emission. 

The other two excesses from June and August 2015, observed during the microquasar outburst of the binary system, also have published spectral analyses available. The June 2015 excess described by \textit{Loh\,16} had a maximum flux of $1.4 \pm 0.5 \times 10^{-6} \, \mathrm{photons \, cm^{-2} \, s^{-1}}$, and a soft power law spectrum ($\Gamma_{PL} = 3.5 \pm 0.8$). These spectral parameters were derived from the highest TS bin in their light-curve. \textit{Piano\,17} do not report the spectral parameters associated with their \textit{Fermi}-LAT analysis, and assume a power law spectrum of $\Gamma_{PL} = 2.1$ for their AGILE analysis, which is the default for an AGILE source with low photon-statistics or an unknown spectrum \citep{pittori_first_2009}. \textit{Piano\,17} show that the AGILE spectral parameters are consistent with the \textit{Fermi}-LAT spectral parameters from \textit{Loh\,16}.  

The August 2015 excess described by \textit{Xing\,20, Figure 4} has an SED calculated in the same way as the August 2016 excess. A soft power law is fitted ($\Gamma_{PL} = 2.5 \pm 0.4$), although like the August 2016 excess a limited number of bins is used (in this case 2). 

\subsubsection{Comparison of spectral analyses}
All three $\gamma$-ray excesses have large uncertainties compared to that of the blazar which was observed over a much longer period. However, there is overlap between the uncertainties of the photon indices of the putative excesses and the blazar index, although statistics enable only a few energy bins to be used for the spectral fits of the excesses. The spectral similarity between the excesses and the blazar means there is insufficient evidence to determine whether the origin of these excesses is B2023+336 or V404\,Cygni. The similarity does suggest the possibility of source confusion, particularly in the case of the \textit{Xing\,20} August 2015 and 2016 excesses where the photon indices lie closer to that of the blazar with smaller uncertainties than that determined during the June 2015 excess.

\subsection{V404\,Cygni Variability Analysis}
\begin{figure}
    \centering
    \includegraphics[width=240pt]{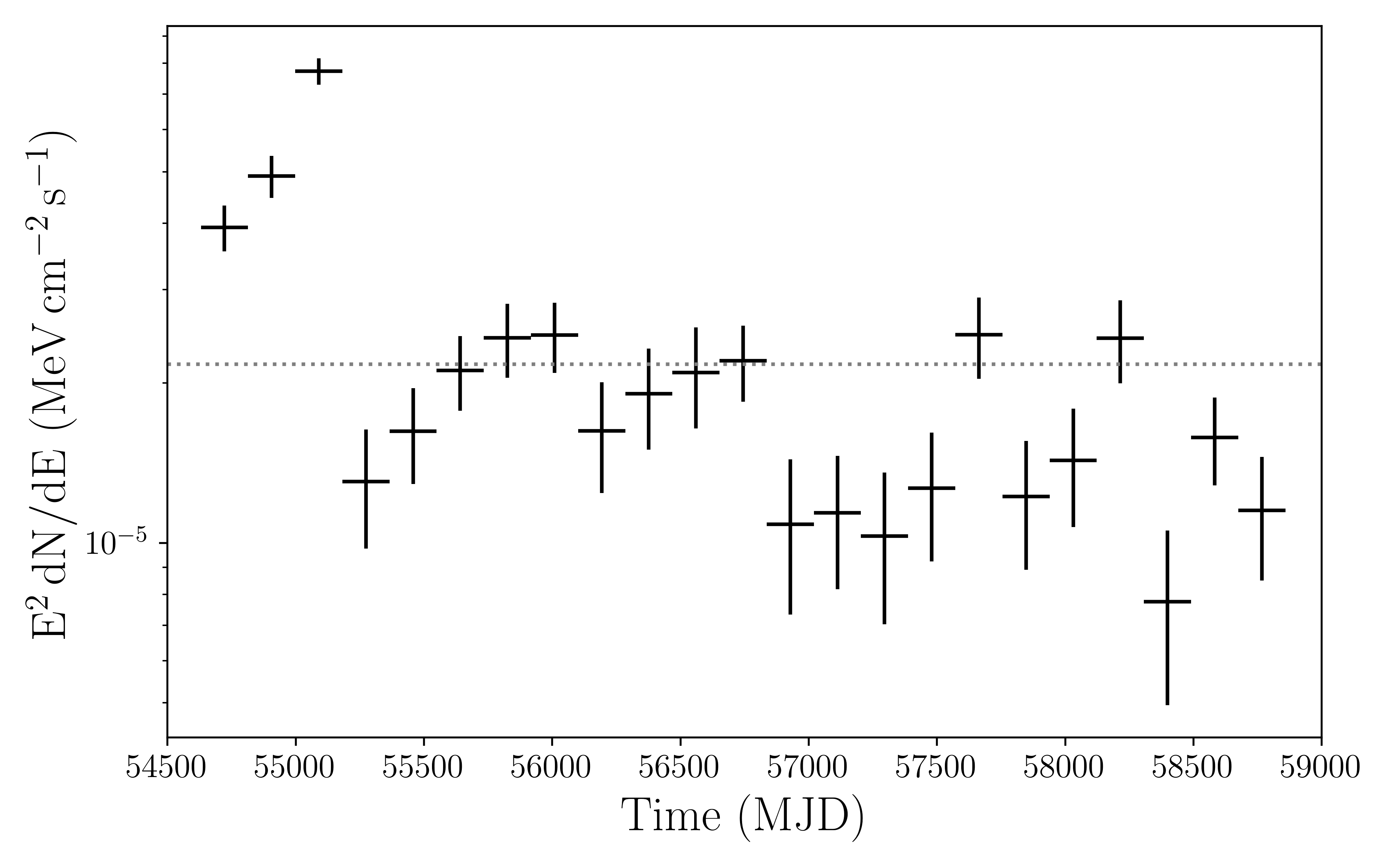}
    \caption{The light-curve of the blazar B2023+336, with approximately 6 month time bins spread out across the observation period given in Table \ref{tbl:params}. The grey dotted line indicates a constant-flux model, which results in a poor fit to the observed data. This is expected given the blazar's 4FGL catalogue variability index ($\mathrm{V.I. = 116}$). }
    \label{fig:6mon_blz}
\end{figure}

As is generally true for FSRQs detected with \textit{Fermi}-LAT \citep{meyer_characterizing_2019}, B2023+336 is variable, with a variability index of 116 \citep{ballet_fermi_2020}. A variability index greater than 72.44 indicates variability on the timescale of months. This long-term variability is illustrated in Figure \ref{fig:6mon_blz}, which shows a rise in flux from the start of the LAT data, with a peak flux of approximately $8 \times 10^{-5} \, \mathrm{MeV \, cm^{-2} \, s^{-1} }$ in early 2010, followed by a sharp drop-off with flux levels between $1$ and $3 \times 10^{-5} \, \mathrm{MeV \, cm^{-2} \, s^{-1} }$ for succeeding bins. From June to December 2015, the flux of B2023+336 plateaus at between $1$ and $2 \times 10^{-5} \, \mathrm{MeV \, cm^{-2} \, s^{-1} }$ with all points within the 95\% confidence limit of one another, indicating a broadly steady $\gamma$-ray flux during the two apparent outbursts of V404\,Cygni.

The three $\gamma$-ray excesses reported from V404\,Cygni are reported on timescales of less than 12 hours rather than months. In general only the brightest sources seen with \textit{Fermi}-LAT have variability that is detectable on short timescales, one example being Cygnus X-3 (\citealt{abdo_modulated_2009}, \citealt{corbel_giant_2012}). For V404\,Cygni to be regarded as a $\gamma$-ray emitter, its emission must reach the $5 \sigma$ level which is conventional for a discovery over this timescale. For reference, the blazar B2023+336 reaches a $5\sigma$ significance over 12 hours in August 2016, with a flux of approximately $5 \times 10^{-4} \mathrm{MeV \, cm^{-2} \, s^{-1}}$. We would expect V404\,Cygni to meet or exceed this flux threshold in order to reach $5\sigma$. 

The June and August 2015 excesses reported by \textit{Loh\,16} and \textit{Xing\,20} respectively, do not reach the $5 \sigma$ level. Furthermore, the fact that the June 2015 excess was identified using a now outdated model and an older catalogue necessitates a repeat analysis of this period with the most recent models. Such an analysis, performed by \textit{Xing\,20}, failed to detect the 2015 events significantly, although as we have noted, differences in the analyses may explain this discrepancy. 

Using the \texttt{Fermipy} light-curve algorithm, we execute a binned light-curve analysis of the \textit{Fermi}-LAT data between June 2015 and September 2015, a time period which covers the outburst of V404 Cygni. For our analysis, we have both background components freed in our model, along with the normalization of all sources within $5 \degree$ of V404\,Cygni's position (including B2023+336). We use a 12 hour independent binning scheme, and place a 95\% confidence upper limit on flux in any bin where the bin TS is less than 4 (corresponding to $2 \sigma$, or $p = 0.05$).

We have established that any potential excess may have a soft spectral energy distribution. The angular resolution of the \textit{Fermi}-LAT is energy dependent, such that the point spread function (PSF) is worse at low energies. This is several degrees in the MeV range, where we expect both the flux of the blazar and also that of the binary to peak. As a measure to test for source confusion between the position of the binary and the blazar, we also execute an identical light-curve, but at the position of the blazar $0.3 \degree$ away.  This will allow us to cross-correlate our results for the position of V404\,Cygni with that of the blazar, to see if any excesses are also seen at the position of the blazar and to perform statistical tests of similarity between the light-curve of the blazar and of V404\,Cygni.

\onecolumn
\begin{figure}
    \centering
    \includegraphics[width=480pt]{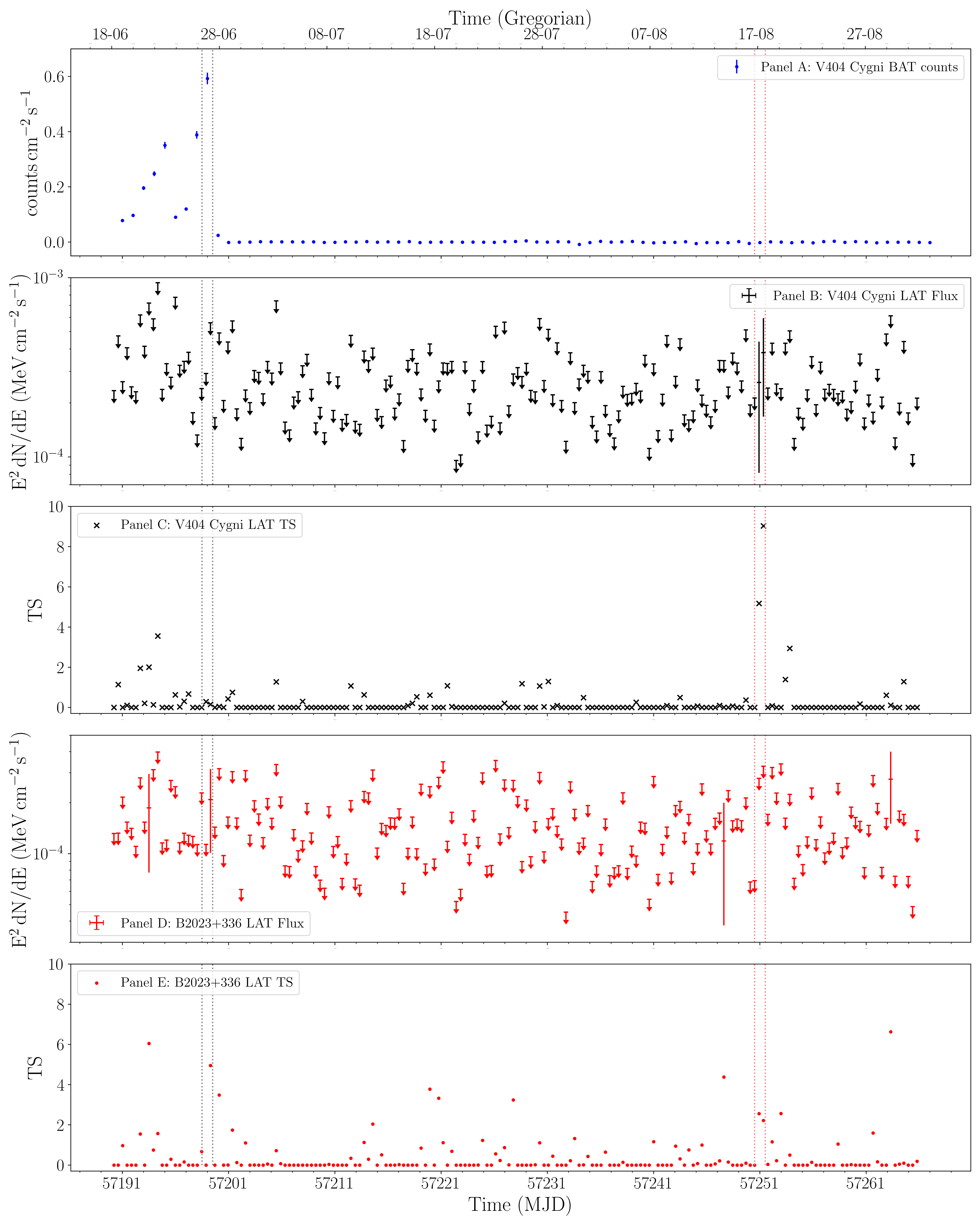}
    \caption{The light-curves of V404\,Cygni and B2023+336 during the 2015 outburst. Panel A shows the \textit{Swift}-BAT light-curve for V404\,Cygni with daily independent binning. Panels B and C show the \textit{Fermi}-LAT light-curve and TS values respectively for this period for V404 Cygni with 12 hour independent binning. Panels D and C show the \textit{Fermi}-LAT light-curve and TS values respectively for B2023+336 with 12 hour independent binning. Units of time are Gregorian, and Modified Julian dates. The vertical grey dotted lines indicate the beginning and end period of the June 2015 excess, whereas the vertical pink dotted lines indicate the beginning and end period of the August 2015 excess.}
    \label{fig:2015_lc}
\end{figure}

\twocolumn

\subsubsection{June 2015 Excess}

\begin{figure}
    \centering
    \includegraphics[width=220pt]{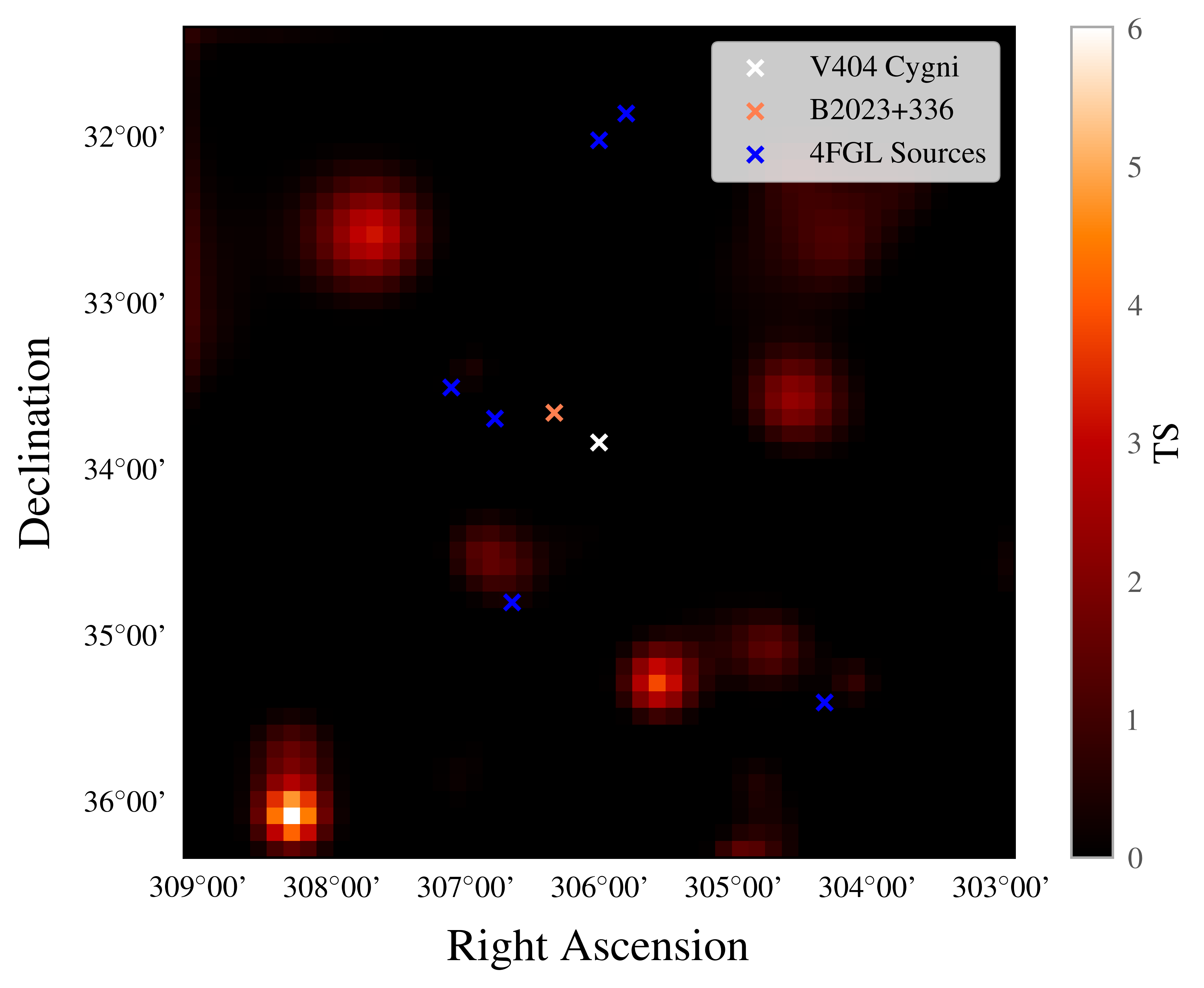}
    \caption{A TS map of the position of V404\,Cygni over the 12 hour period (MJD 57199.25-57199.75) where \textit{Loh\,16} describe their $\gamma$-ray excess (coincident with the peak in X-ray brightness). We observe no excess of $\gamma$-rays from the position of V404\,Cygni during this time that cannot be accounted for by any of the neighbouring 4FGL sources, or the background models. This TS map is taken over the full effective \textit{Fermi}-LAT energy range of $100 \, \mathrm{MeV}$ to $300 \, \mathrm{GeV}$ with $0.1 \degree$ spatial bins.}
    \label{fig:June15TS}
\end{figure}

 Figure \ref{fig:2015_lc} shows a comparison between the V404\,Cygni light-curve (black) and B2023+336 (red) $\gamma$-ray flux and the TS value of each bin during the June 2015 outburst period, as well as the \textit{Swift}-BAT light-curve of V404\,Cygni for this time. We do not see any $\gamma$-ray excess from the position of V404\,Cygni during June 2015, when \textit{Loh\,16} report a $4 \sigma$ excess at the peak of the \textit{Swift}-BAT light-curve highlighted by the TS map in Figure \ref{fig:June15TS}. This is not entirely surprising, as both \textit{Loh\,16} and \textit{Piano\,17} used older background models and an older catalogue  for their analysis. A key difference between the 3FGL used by \textit{Loh\,16} and the 4FGL used in our analysis (and \textit{Xing\,20}) is the addition of weighting in the maximum likelihood method employed in LAT analysis. The weighted maximum likelihood method better reflects the systematic uncertainties of the instrument, and results in larger parameter uncertainties and correspondingly smaller TS values. This could explain why an apparently significant time bin in the \textit{Fermi}-LAT results of \textit{Loh\,16} and \textit{Piano\,17} is no longer seen when using the 4FGL, although this does not explain the AGILE result described in \textit{Piano\,17}. This result is in agreement with \textit{Xing\,20}, who similarly find no $4 \sigma$ $\gamma$-ray excess in June, although there is a lack of information regarding the analysis parameters of \textit{Xing\,20.}
 
A further difference between our analysis and that of \textit{Loh\,16} is that \textit{Loh\,16} use an unbinned analysis. A binned analysis is preferred for sources on the Galactic plane; however, we also performed an unbinned analysis over the 12 hour period shown in Figure \ref{fig:June15TS}, and find that this analysis agrees with the binned result.

\subsubsection{August 2015 Excess}

\textit{Xing\,20} report a separate $4 \sigma$ $\gamma$-ray excess in August 2015 through use of independent binning. Two bins in our V404\,Cygni light-curve have significances above the upper limit threshold in August 2015, ($\sim$ MJD 57521), which is the same time period as that reported by \textit{Xing\,20} in their analysis. Whilst \textit{Xing\,20} reports this excess at the $4 \sigma$ level, we find that one bin reaches the $2 \sigma$ level, and the second, consecutive, bin reaches $3 \sigma$, with no corresponding rise in the count rate of the \textit{Swift}-BAT light-curve. We do not see a corresponding flux increase in the light-curve of the blazar. Given that we use the same LAT catalogue and background models as \textit{Xing\,20} it is likely that this discrepancy in results is down to the photon selection (we use energies greater than 100 MeV, and \textit{Xing\,20} use energies greater than 300 MeV), and potentially other differences between our analysis and that of \textit{Xing\,20}.

With $2 \sigma$ and $3 \sigma$ consecutive bins over such a long time period we must consider the likelihood of an apparently significant result arising simply by chance. Out of 184 bins in our light-curve of V404\,Cygni, we find 2 bins with at least a $2 \sigma$ result. Looking at the $3 \sigma$ bin in particular, there is a 1 in 370 chance that this is a statistical anomaly. Considering that we have 184 separate bins, it is important to quantify the probability that this result arises by chance, as the number of bins is comparable to the chance probability. 

The binomial distribution provides a suitable representation of our bins, and we calculate that the chances of finding exactly one $3 \sigma$ bin out of 184 to be 30.3\%, with a probability of finding at least one $3 \sigma$ result rising to 39.2\%. We therefore do not believe this August 2015 excess to be a significant $\gamma$-ray flare. 

\subsubsection{August 2016 Excess}

\begin{figure}
    \centering
    \includegraphics[width=480pt]{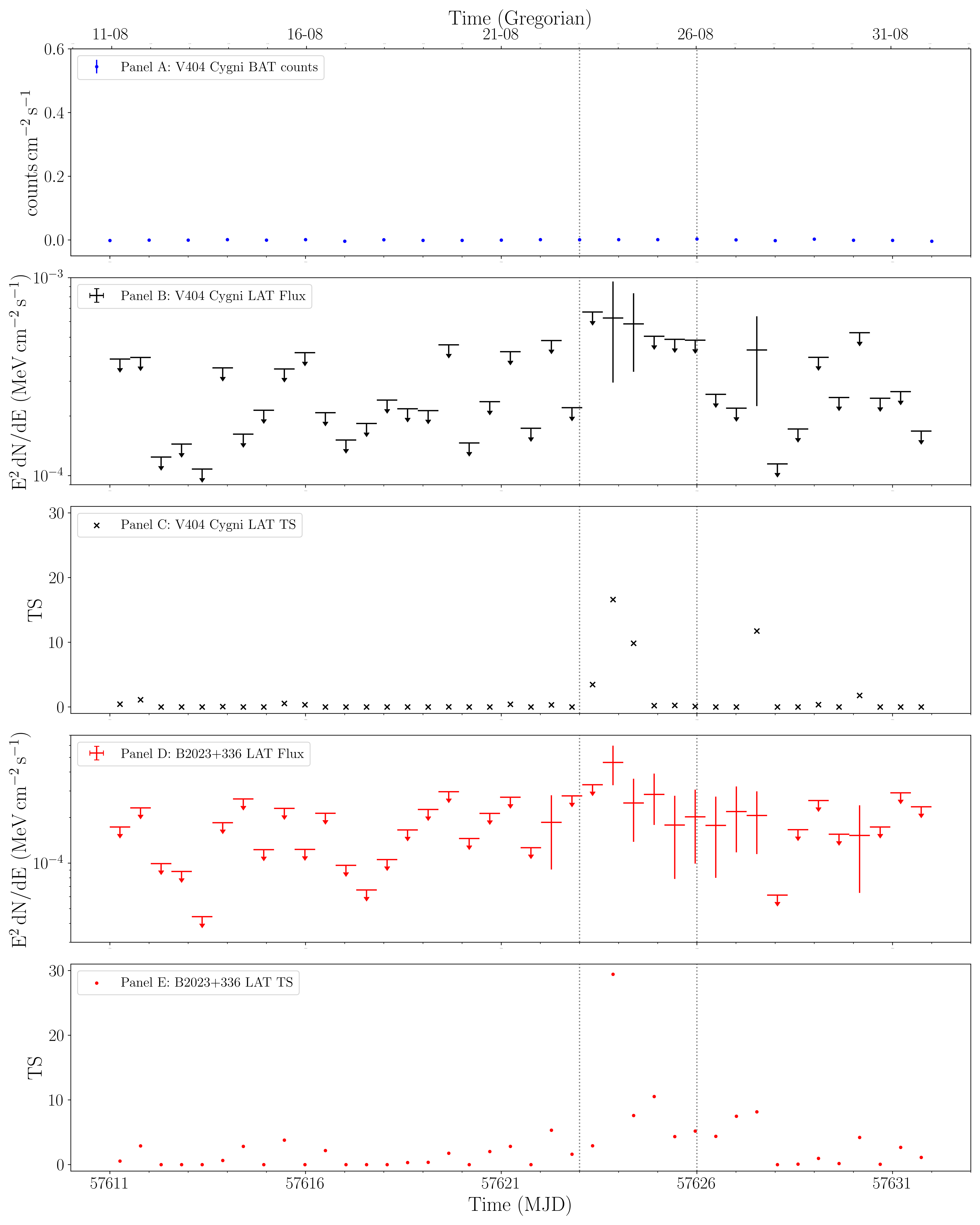}
    \caption{The light-curves of V404\,Cygni and B2023+336 during August 2016, when \textit{Xing\,20} claim their detection of V404\,Cygni. Panel A shows the \textit{Swift}-BAT light-curve for V404\,Cygni with daily independent binning. Panels B and C show the \textit{Fermi}-LAT light-curve and TS values respectively for this period for V404 Cygni with 12 hour independent binning. Panels D and E show the \textit{Fermi}-LAT light-curve and TS values respectively for B2023+336 with 12 hour independent binning. Units of time are Gregorian, and Modified Julian Dates. The vertical grey dotted lines indicate the beginning and end of the 3-day time bin \textit{Xing\,20} used to identify the August 2016 excess. We also observe $\gamma$-ray emission outside this time period at comparable TS values. }
    \label{fig:2016_lc}
\end{figure}

\textit{Xing\,20} also claims a more significant $7 \sigma$ $\gamma$-ray excess from V404\,Cygni during August 2016: a year after the outburst finishes. This is by far the most significant excess in their light-curve. Figure \ref{fig:2016_lc} shows the light-curves of V404\,Cygni and B2023+336 during August 2016, around the result claimed by \textit{Xing\,20}. Using independent 12 hour binning, we do not find as high TS values as \textit{Xing\,20} (although, as for the analysis of the August 2015 event, our photon selection and analysis parameters differ to those of \textit{Xing\,20}). Nevertheless, we do see three bins at the $3 \sigma$ to $4 \sigma$ level over a short period, with measurable fluxes. However, when we look at the light-curve of B2023+336, we also detect fluxes in these bins and others around this time. As $\gamma$-ray fluxes are detected at both the binary and blazar simultaneously, this suggests confusion as to whether the flare originates from V404\,Cygni or B2023+336.

There are no available multi-wavelength observations of B2023+336 during the time of this detection. However, neither the optical AAVSO light-curve or X-ray \textit{Swift}-BAT light-curve of V404\,Cygni (Figure \ref{fig:2016_lc}) show any enhancement in their respective wavebands during August 2016, nor is there any significant enhancement since the December 2015 flare. For V404\,Cygni to form a jet which emitted $\gamma$-rays, we would expect an enhancement in the X-ray flux, similar to that seen in June 2015 (Figure \ref{fig:2015_lc}). As we do not see this, this suggests that there was no outburst from V404\,Cygni at this time, supporting the hypothesis that the flare is from the blazar, not the binary.

\onecolumn
\twocolumn

\subsubsection{Statistical Tests of Similarity}
\label{sec:similarity}
In order to look for similarity between the $\gamma$-ray emission of V404\,Cygni and the blazar, B2023+336, we employ a 2-sample Kolmogorov-Smirnov (KS) test (\citealt{kolmogorov_sulla_1933} \citealt{smirnov_table_1948}) in order to explore the hypothesis of source confusion. The 2-sample KS test essentially tests whether two numerical distributions are drawn from some common overall distribution, by calculating a KS statistic using Equation \ref{eqn:KS}.

\begin{equation}
    D_{a, \: b} = \sup |F_{1, \: a(x)} - F_{2, \: b(x)}|
    \label{eqn:KS}
\end{equation}

\begin{figure}
    \centering
    \includegraphics[width=240pt]{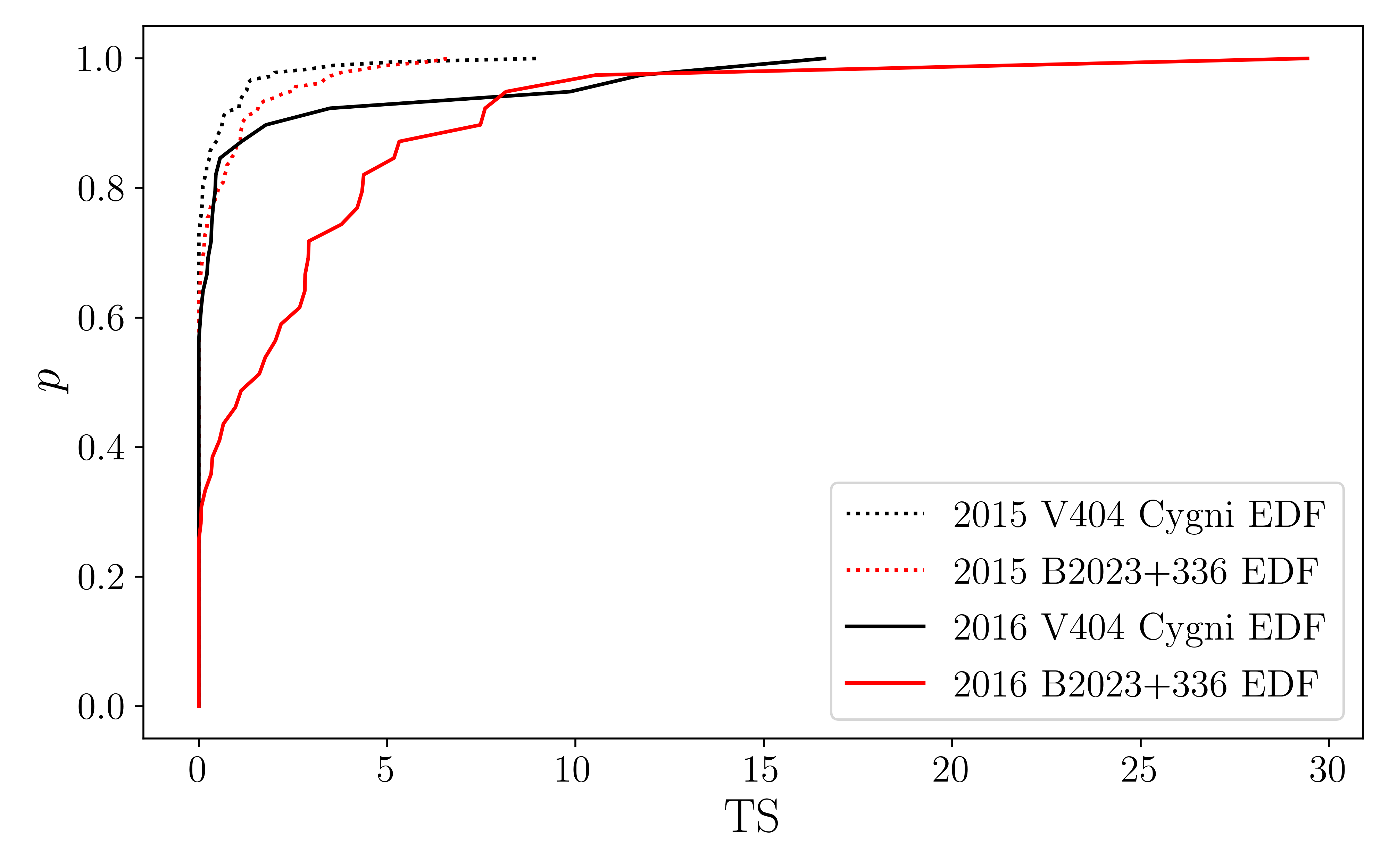}
    \caption{The empirical distribution functions of V404 Cygni and B2023+336 during the 2015 outburst and the 2016 flare for the TS of each time bin. The B2023+336 and V404\,Cygni distributions in 2015 are very similar, and are dominated by the noisy background. During 2016, the V404 Cygni distribution reaches a higher peak TS, with a higher probability of increased TS values over 2015. The most significant emission comes from B2023+336 during 2016, where we see a TS peak higher than the V404 Cygni distribution.}
    \label{fig:CDF}
\end{figure}

Here, $D_{a,\: b}$ is the KS statistic for two samples $a(x)$ and $b(x)$, where the KS statistic is equal to the supremum of the absolute difference between the empirical distribution functions (EDFs) of the two samples. The EDFs for all of our samples are shown in Figure \ref{fig:CDF}. Alternative tests of similarity exist, such as the Mann-Whitney test \citep{mann_test_1947}, or the well known Student's t-test. However, we use the KS test as it is more powerful in detecting changes in the shape of the distribution, which is essential when analysing time series data. 

For the 2015 outburst period, we find $D = 0.11$ indicating a $p$-value of $p = 0.23$ for the hypothesis that the samples are drawn from separate distributions. This is unsurprising, as Figure \ref{fig:CDF} shows that the EDFs for the blazar and binary at this time are not substantially different and although B2023+336 is a luminous $\gamma$-ray source, it is not known to be regularly detected on timescales as short as 12 hours. Both sources are likely both dominated by the same noisy $\gamma$-ray background on the Galactic plane during the outburst, which provides the most likely source of the common distribution of TS values during the 2015 outburst. 

For the 2016 outburst period, we find $D = 0.425$ corresponding to $p = 0.001$ for the same hypothesis, indicating a significant difference in the TS distributions of both the blazar and binary system. In Figure \ref{fig:CDF} we see an increased probability of higher TS values for both systems when compared to the 2015 outburst period where both systems are noise dominated, indicating increased $\gamma$-ray emission from both B2023+336 and the position of V404\,Cygni. The plots of TS against time shown in Figure \ref{fig:2016_lc} for August 2016 also show that the increased TS (and therefore flux) occurs for both systems at the same time, with the peak of both light-curves occurring in the same bin indicating that this emission could have the same origin.  Given that the EDF of B2023+336 shows an increased probability of higher TS values, and therefore more significant $\gamma$-ray emission than from V404\,Cygni, this serves as statistical evidence at the $3 \sigma$ level that the origin of the August 2016 flare is B2023+336 rather than V404 Cygni. 

In conjunction with the lack of X-ray emission observed by \textit{Swift}-BAT from V404\,Cygni during August 2016, we believe that there is sufficient evidence to state that the August 2016 $\gamma$-ray flare originates from B2023+336 and not V404\,Cygni. Any $\gamma$-ray emission observed from the position of V404\,Cygni is a product of source confusion with B2023+336, due to the properties of the LAT itself (resolution, point spread function etc), which are less precise at the lower energies where this flare occurs, as established in Section \ref{sec:spectra}. Given that the LMXB system was in quiescence at this time, a blazar origin for this emission is much more likely. 

\subsubsection{Finding significant bins by chance}
\textit{Xing\,20} find the August 2015 and August 2016 excesses by running a light-curve over 11.5 years with 3-day independent binning, having also created but discarded light-curves using 1-day and 6.5-day binning. This light-curve is shown in \textit{Xing\,20 Figure\,3}. 

In addition to the $\gamma$-ray excesses described above, which we attribute to the large number of bins searched and source confusion with a B2023+336 flare, they find 10 other 3-day bins with TS values at the $3 \sigma$ level or above. The first four of these excesses (in time), labelled Period 1 in \textit{Xing\,20 Figure\,3}, occur during the first 18 months of the \textit{Fermi}-LAT mission. From our mission-long light-curve of B2023+336, we can see that for the first 18 months of the \textit{Fermi}-LAT mission, the flux is in an enhanced state with respect to all later bins. Both \textit{Loh\,16} and the \textit{Fermi} All-Sky Variability Analysis team \citep{ackermann_fermi_2013} also report on the enhanced state of the blazar. As Period 1 overlaps with the enhanced flux state of the blazar, and considering our evidence with regards to source confusion, particularly during flares, we believe the $\gamma$-ray flux enhancement during Period 1 to be from the blazar. 

Having accounted for the flux excesses in Period 1 and the August 2015 and 2016 excesses, we note that 6 other $\gamma$-ray excesses are described in \textit{Xing\,20 Figure\,3}. As this light-curve covers 11.5 years, with 3 day binning, we estimate there to be approximately 1400 bins in this time. We are able to employ the binomial distribution to predict how many $\gamma$-ray excesses we are likely to occur at the $3 \sigma$ level.  We find that there is an 9.25\% chance of finding these 6 $\gamma$-ray excesses by chance, indicating a strong possibility that there is no $\gamma$-ray emission from V404\,Cygni shown in the \textit{Xing\,20 Figure\,3} light-curve that cannot be accounted for with either source confusion with B2023+336 or by considering the effect of apparently significant bins arising by chance.

\section{Discussion}
Using recent \textit{Fermi}-LAT background models and instrument response functions, together with the 4FGL, we analyse the position of V404\,Cygni in the Pass 8 \textit{Fermi}-LAT data. Previous works \textit{Loh\,16}, \textit{Piano,17} and \textit{Xing\,20} have identified 3 $\gamma$-ray excesses over the course of the \textit{Fermi}-LAT mission which could be indicative of high energy $\gamma$-ray emission from V404\,Cygni.

The first of these excesses is described by \textit{Loh\,16} and \textit{Piano\,17}, and occurred in June 2015 coincident with the hard X-ray peak of the outburst, during the peak of the AAVSO optical light-curve, and shortly following the peak radio emission. The background models, catalogue and instrument response functions used in this analysis are now superseded by more accurate models, and when we carry out our own binned (and unbinned) analyses, we find no significant $\gamma$-ray emission. We believe this $\gamma$-ray excess to be a product of the older models available at the time for \textit{Fermi}-LAT data analysis, rather than a statistically significant detection. This is supported by the fact that the peak significance only reaches $4 \sigma$ in the \textit{Loh\,16 analysis}, and not the conventional $5 \sigma$ level required to claim a detection, although a slightly lower significance may be acceptable in light of multi-wavelength data supported by theory. Whilst there appears to be no significant excess from V404\,Cygni as seen with \textit{Fermi}-LAT based on our analyses, this does not discount the excess observed within the AGILE data discussed in \textit{Piano\,17}, which remains the strongest independent evidence for $\gamma$-ray production in V404\,Cygni.

The next of these excesses is a separate $4 \sigma$ excess at the end of the outburst in August 2015, claimed by \textit{Xing\,20}. Unlike the first, there is no corresponding X-ray enhancement, but we do also see this excess in our own analysis. However, a wider issue with apparently significant bins arising by chance both in our own light-curve (Figure \ref{fig:2015_lc}) and the work of \textit{Xing\,20} leads us to believe that this is probably a chance occurrence. 

The final, and most significant, claim of $\gamma$-ray emission occurring from V404\,Cygni was the $7 \sigma$ August 2016 excess reported by \textit{Xing\,20}, after the X-ray outburst had finished. We find that this is more than likely a product of source confusion with B2023+336, which also appeared to be active at this time. Given that the $\gamma$-ray emission from B2023+336 is present longer, more consistently, and more significantly than that reported from the position of V404\,Cygni, and the spectral similarity between this excess and the spectrum of B2023+336, we believe that this $\gamma$-ray excess is a product of source confusion between the blazar and V404\,Cygni. This is supported by the fact that V404\,Cygni was in quiescence at this time, and is not likely to become an active microquasar again for approximately another decade or two.

A definitive identification of $\gamma$-ray emission from a new binary system would be an important result, as so few are detected with \textit{Fermi}-LAT. An interesting prospect in this respect is AMEGO \citep{mcenery_all-sky_2019}, proposed to launch in 2030. As AMEGO will operate in the MeV range, where we would expect V404\,Cygni to emit $\gamma$-rays, and will operate with greater sensitivity and resolution than \textit{Fermi}-LAT. It is very possible that, if V404\,Cygni becomes active again, AMEGO will make a significant detection. The 2030 target launch date, and the 1 to 2 decade microquasar cycle of V404\,Cygni, means the next time this system becomes active the scientific community will hopefully have a clearer picture as to the high energy physics involved in such a system. 
\label{dis}

\section*{Data Availability}
All data used within this paper is from the NASA \textit{Swift} and \textit{Fermi} missions. It is publicly available for download from the relevant servers on the NASA website. The \textit{Fermi}-LAT data analysis tools are also publicly available from the same site.  

\section*{Acknowledgements}

The authors would like to acknowledge the excellent data and analysis tools provided by the NASA \textit{Fermi} collaboration, without which this work could not be done.  In addition, this work has made use of the SIMBAD database, operated at CDS, Strasbourg, France and Montage, funded by the National Science Foundation.  We would also like to thank Chris Done, Matthew Capewell and Jamie Holder for useful discussions. We thank the referee for constructive feedback in producing this work. 

MH acknowledges funding from Science and Technology Facilities Council (STFC) PhD Studentship ST/S505365/1, and PMC and CBR acknowledge funding from STFC consolidated grant ST/P000541/1. 




\bibliographystyle{mnras}
\bibliography{references.bib} 




\bsp	
\label{lastpage}
\end{document}